\documentclass[pra,twocolumn,showpacs,preprintnumbers,amsmath,amssymb]{revtex4}
\usepackage{graphicx}
\usepackage{dcolumn}
\usepackage{bm}
\usepackage{latexsym,epsfig}

\begin{document}

\preprint{preprint}
\title{Phase separated charge density wave phase in two species extended Bose-Hubbard model}
\author{Tapan Mishra}
\email{tapan@iiap.res.in} \affiliation{ Indian Institute of
Astrophysics, II Block, Kormangala, Bangalore, 560 034, India.}
\author{B. K. Sahoo}
\email{B.K.Sahoo@rug.nl} \affiliation{ KVI, University of Groningen,
NL-9747 AA Groningen, The Nederlands}
\author{Ramesh V. Pai}
\email{rvpai@unigoa.ac.in} \affiliation{ Department of Physics, Goa
University, Taleigao Plateau, Goa 403 206, India. }

\date{\today}

\begin{abstract}
We study the quantum phase transitions in a two component
bose-mixture in a one-dimensional optical lattice. The calculations
have been performed in the framework of the extended Bose-Hubbard
model using the finite size density matrix renormalization group
method.  We obtain different phase transitions for the system for
integer filling. When the inter-species on-site and the nearest
neighbor interactions are larger than the intra-species on-site and
also the nearest neighbor interactions, the system exhibits a phase
separated charge density wave (PSCDW) order that is characterized by
the two species being spatially separated and existing in the
density wave phases.
\end{abstract}


\maketitle

\section{Introduction}
Ultra cold atoms in the optical lattices can provide new insights
into quantum phase transitions~\cite{review}. The remarkable control
of the interaction strengths between the atoms by tuning the laser
intensity ~\cite{greiner} leads to the experimental realization of
the superfluid (SF) to Mott insulator (MI) transition which was
predicted by Jaksch \textit{et al} ~\cite{jaksch}. The observation
of the SF to MI transition in the one-dimensional (1D) optical
lattice ~\cite{stoferle} has further enhanced the interest in the
search for new quantum phases in the low dimensional bosonic
systems.  Recent realization of Bose-Einstein condensation (BEC) in
strongly dipolar $^{52}Cr$ atoms ~\cite{pfau} has enlarged the
domain of interaction space to investigate various quantum phase
transitions and other possible subtle characters of bosons at
different limit that can be experimentally observed.  When atoms
with large dipole moments loaded into the optical lattices, the long
range interaction between the atoms plays very important role, in
addition to the onsite interaction, in the determination of the
ground state. This system can be described by the extended
Bose-Hubbard (EBH) model, which includes the nearest neighbor
interaction along with the onsite repulsion, exhibiting many new
phases such as charge density wave (CDW) (sometime known as mass
density wave (MDW)) \cite{pai,white}, Haldane insulator
order\cite{torre} and exotic supersolid ~\cite{batrouni} et cetera..

On the other hand, the study of mixtures of atoms such as bose-bose
\cite{ref1,ref2}, bose-fermi \cite{ref3,ref4} and fermi-fermi
\cite{ref5,ref6} have attracted much attention in recent years
because of the successful realization of such systems in optical
lattices. In the  case of bose-bose mixture, the theoretical models
take on-site intra- and inter-species interactions into
consideration  to describe the system  in a large domain of system
parameters and the competition between them opens-up many new
possible quantum phases ~\cite{kuklov,demler,sengupta,mathey}.
Recent studies in the one-dimensional two species bose mixtures have
revealed a spatially phase separated (PS)
phase~\cite{mishra,cederbaum}, when the inter-species interaction is
greater than the intra-species interaction. This phase separation
can either of SF or MI type depending upon the strong interplay
between the on-site intra-species and inter-species
interactions~\cite{mishra}. In this context, it is very interesting
and relevant to study the bose mixtures of dipolar atoms to
investigate the underlying influence of long range interactions on
these phases. Prior theoretical studies of such systems will be
helpful to guide the direction of experimental investigations. Our
aim of this work is to extend the search for new possible phases by
taking into account the nearest neighbor interactions along with the
onsite intra- and inter- species interactions in the two species
bose mixture which we have studied earlier~\cite{mishra}. We employ
the finite size density matrix renormalization group (FS-DMRG)
method to study the system.

We have organized the remaining part of the paper in the following
way. In Sec. II, we present the theoretical model that we have
considered, followed by  the method of calculations. We have given a
brief discussions of the cases that we have taken into account in
this work and a detailed analysis of the results in Sec. III and
Sec. IV, respectively. Finally, we conclude our findings in the last
section.

\section{Model Hamiltonian and method of calculations}
\label{sec:model} In this work, we consider bose mixtures of dipolar
atoms in an 1D optical lattice. The corresponding effective
Hamiltonian for such systems can be expressed as
\begin{widetext}
\begin{eqnarray}\label{eq:ham}
\mathcal{H} &=&\sum_{c=a,b} \left \{ \sum_i \frac{U}{2}^{c}
n_{i}^{c}(n_{i}^{c}-1) +  \sum_{<i,j>} \left[ -t^{c}(c_{i}^{\dagger}
c_{j}+h.c.)
 + V^{c} n_{i}^{c}n_{j}^{c} \right] \right \}+ \sum_i U^{ab} n_{i}^{a}n_{i}^{b}
+V^{ab}\sum_{<i,j> }n_{i}^{a}n_{j}^{b},
\end{eqnarray}
\end{widetext}
where $c_i$ and $c_i^{\dagger}$ (with $c=a,b$) represent the
annihilation and creation operators, respectively, for bosonic atoms
of $a$ or $b$ types on site $i$ whose number operators are defined
by $n^c_i=c^\dagger_ic_i$. In the above equation, $t^{c}$, $U^c$ and
$V^c$ are the amplitudes for the hopping between nearest neighboring
sites, the on-site and nearest neighbor intra-species repulsive
interactions, respectively. The inter-species on-site and nearest
neighbor interactions are represented by $U^{ab}$ and $V^{ab}$,
respectively. It is obvious from Eq. (\ref{eq:ham}) that there are
at least 8 independent parameters in the model. Since it is not
possible to vary all these parameters at a time to grasp the
underlying physics of the above model, we  restrict ourselves to
some special range of parameters which are guided by some cases that
have already been studied earlier \cite{pai,mishra}.  We also keep
the symmetry between both the $a$ and $b$ types of bosons by
assuming $t^a=t^b=t$, $U^a=U^b=U$ and $V^a=V^b=V$. We  scale the
energy of the whole system with respect to $t$ by setting its value
as unity.

In our earlier study in the absence of nearest neighbor
interactions; i.e. $V=V^{ab}=0$, many interesting phases had been
predicted. In particular, our work revealed the possible existence
of both the species being in SF phases, the system as a whole
existing as a MI and phase separated superfluid (PSSF) and phase
separated Mott insulator (PSMI)~\cite{mishra} by varying the onsite
interaction strengths of both $a$ and $b$ type bosons. It was shown
that a phase separation between SF phases of $a$ and $b$ is possible
when $U^{ab}$ is considered (slightly) larger than $U$. When the
total density of the system was an integer ($\rho=1$) with density
of each species equal to half ($\rho_a=\rho_b=1/2$), we had
predicted SF, PSSF and PSMI phases in the $U$ and $U^{ab}$ phase
space. Furthermore, in the incommensurate densities with $\rho_a=1$,
$\rho_b=1/2$ and $\rho=3/2$, we had found only the SF and PSSF
phases. In contrast to this case, when $U^{ab} \le U$ was
considered, only the SF phase was possible for the incommensurate
densities while signatures of both the SF and MI phases with a
continuous SF to MI phase transitions were found for the
commensurate densities.

The aim of this work is to investigate how these phases evolve in
the presence of intra- and inter-species nearest neighbor
interactions. For a better analysis of a particular situation, we
restrict ourselves to the commensurate densities, especially the
case when $\rho_a=\rho_b=1/2$ with $\rho=1$. This choice is governed
by the knowledge that we have acquired from the following studies in
the phase diagram of (i) the extended Bose Hubbard model for density
$\rho=1$ ~\cite{pai} for a single species boson and (ii) the two
species Bose Hubbard model for densities $\rho_a=\rho_b=1/2$ and
$\rho=1$ ~\cite{mishra}. Our analysis of the results from the
present study is based upon the findings of the above two cases and
conclusions are drawn with respect to them.

Model (\ref{eq:ham}) is a difficult problem to study analytically.
We have employed FS-DMRG method with open-boundary condition to
determine the ground state. This method has been proved to be one of
the most powerful techniques for 1D systems
\cite{whiteprl,pai,dmrgreview}. We have considered a soft-core case
by keeping the number of bosonic states per site for each species as
four ($4$). We allow up to $128$ states in the density matrix of the
left and right blocks in each iteration of the FS-DMRG calculations.
The weight of the states neglected in the density matrix of left and
right blocks are less than $10^{-6}$. To get a better convergence of
the ground state energies of various phases, especially for larger
values of intra- and inter-species nearest neighbor interactions, we
have performed the finite size sweeping procedure ~\cite{pai} twice
in each iteration of the FS-DMRG method.

To identify the ground states of various phases of the model
Hamiltonian given by Eq. (\ref{eq:ham}), we calculate the single
particle excitation gap $G_L$ defined as the difference between the
energies needed to add and remove one atom from a system of atoms;
i.e.
\begin{equation}
G_L=E_L(N_a+1,N_b)+E_L(N_a-1,N_b)-2 E_L(N_a,N_b) \label{eq:gap}
\end{equation}
We also calculate the on-site number density as
\begin{equation}\label{eq:ni}
    \langle n^c_i \rangle=\langle
\psi_{LN_aN_b}| n^c_i |\psi_{LN_aN_b}\rangle.
\end{equation}
Here $c$, as mentioned before, is an index representing type $a$ or
$b$ bosons, with $N_a$ ($N_b$) corresponds to total number of $a$
($b$) bosons in the ground state $|\psi_{LN_aN_b}\rangle$ of a
system of length $L$ with the  ground state energy $E_L(N_a,N_b)$.

In $1D$ the appearance the SF phase is signaled by $G_L \rightarrow
0$ for $L\rightarrow \infty$. However, for a finite system $G_L$ is
finite, and we must extrapolate to $L\rightarrow \infty$ limit,
which is best done by finite size scaling of gap\cite{pai,paiprl}.
In the critical region
\begin{eqnarray} G_L \equiv L^{-1} f(L/\xi)
\end{eqnarray}
where $\xi$ is the correlation length which diverges in the SF
phase. Thus plots of $LG_L$ versus interaction for different values
of $L$ coalesce in the SF phase. On the other hand, when this trend
does not follow then the system can be said to be in the MI phase.

We also define the CDW order parameter for the bosons as
\begin{equation}
O^{c}_{CDW}=\frac{1}{L}\sum_i \langle\psi_{LN_aN_b}| (|n^c_i -
\rho|)|\psi_{LN_aN_b} \rangle.
\end{equation}
So when the CDW order parameter of the system is finite then the
system is assumed to be in the CDW phase. Since $\rho$ of the system
is constant, it is clear
 from the above equation that the density of the bosons will oscillate when
they are in the CDW phase.

To find whether the ground state is in the spatially phase
separated, we calculate the PS order parameter, which is given by
\begin{equation}
O_{PS}=\frac{1}{L}\sum_i \langle\psi_{LN_aN_b}| (|n^a_i - n^b_i|)
|\psi_{LN_aN_b} \rangle.
\end{equation}

When $O_{PS}$ is finite, the system is said to be in the PS phase. Therefore,
the system can be simultaneously in PS and one of the SF, MI or CDW phases
which can be distinguished by determining both $O_{PS}$ and one of the above
properties to identify the other corresponding phase.

\begin{figure}[htbp]
  \centering
  \epsfig{file=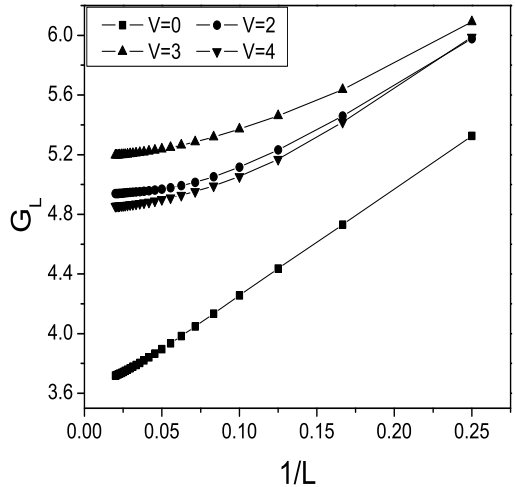,width=8cm,height=8cm}
  \caption{Gap $G_L$ versus $1/L$ for different values of $V$ for $U=9$, $\Delta U=1.05$ and
  $\Delta V=0.5$. $G_{L\rightarrow \infty}$ converges to a finite
  values signaling Mott insulator phase.}
     \label{fig:fig1}
\end{figure}

\begin{figure}[htbp]
  \centering
  \epsfig{file=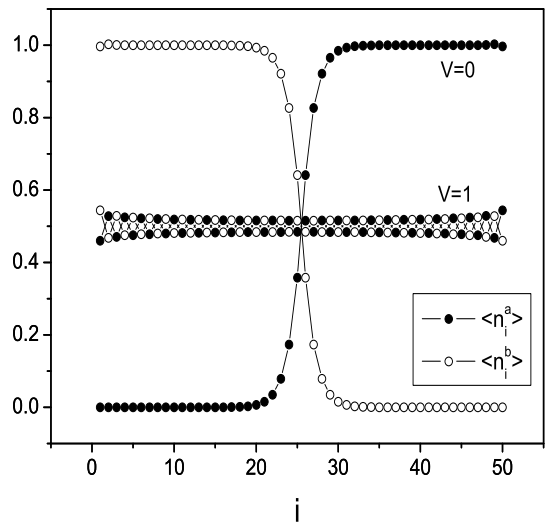,width=8cm,height=8cm}
  \caption{Plots of $\langle n^a_i\rangle$ and  $\langle n^b_i\rangle$ versus $i$ for
  $V=0$ and $1$, respectively, showing PSMI and MI phases.}
     \label{fig:fig2}
\end{figure}

\begin{figure}[htbp]
  \centering
  \epsfig{file=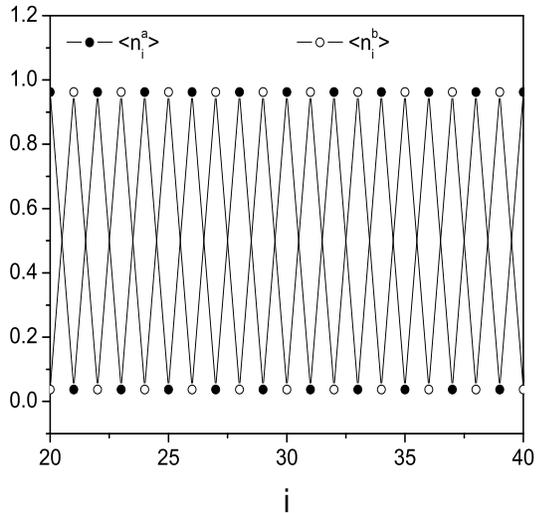,width=8cm,height=8cm}
  \caption{Plots of $\langle n^a_i\rangle$ and  $\langle n^b_i\rangle$ versus $i$ for
  $V=4$ showing intermingled CDW phases for $a$ and $b$ types of bosons.}
     \label{fig:fig3}
\end{figure}

\section{Pre-analysis of results}

Before presenting the details of our results, we first summarize the
main features of our study here. In this study, our main focus is to
understand the effects of intra- and inter-species nearest neighbor
interactions between the atoms on the PSMI phase. As mentioned
earlier, the PSMI phase is possible only if $\Delta U \equiv
U^{ab}/U >1$, when $V=V^{ab}=0$. As we show below there is a
stringent condition for the PSMI phases when the nearest neighbor
interactions are finite. In the present work, we fix $\Delta U=1.05$
and consider two values of intra-species on-site interaction $U=6$
and $9$. Our previous study~\cite{mishra} had yielded the ground
state of model (\ref{eq:ham}) with $\rho_a=\rho_b=1/2$ is in PSMI
phase for these values of intra- and inter-species on-site
interactions. Similarly, the phase diagram of the single species EBH
model~\cite{pai} shows that ground state for $U=6$ varies first from
MI to SF as the nearest neighbor interaction $V$ increases from zero
and then to the CDW phase for the larger values of $V$. However, for
$U=9$, there is no SF phase sandwiched between the MI and CDW phases
and the transition between them is direct. We present below the
results obtained from this investigation, where the nearest neighbor
interactions are finite.

\begin{figure}[htbp]
  \centering
  \epsfig{file=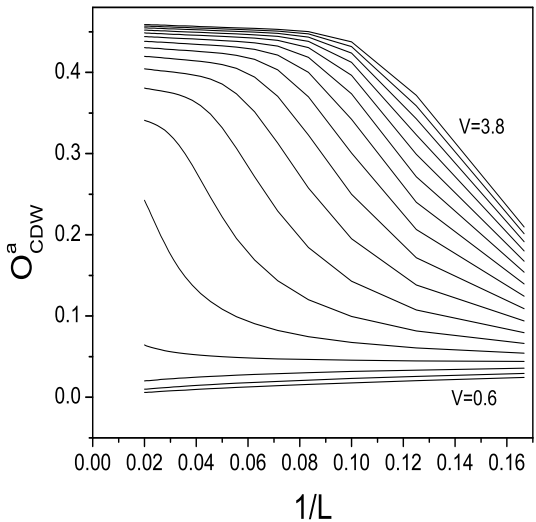,width=8cm,height=8cm}
  \caption{Plot of CDW order parameter $O^a_{CDW}$ for $a$-type atoms as a function of $1/L$
for values of $V$ ranging from $0.6$ to $3.8$ in steps of $0.2$. The
$O^a_{CDW}$ goes to zero for $V<V_C\simeq1.2$ where as it is nonzero
for higher values of $V$ which shows the transition to CDW phase at
$V_C\simeq1.2$.}
     \label{fig:fig4}
\end{figure}
One feature which emerges from our study is that when intra- and
inter-species nearest neighbor interactions are finite, the PSMI
phase is possible only for $V^{ab} > V$. We find that for a fixed
$\Delta V=V^{ab}/V=1.25$ and $U=6$, the ground state evolves from
PSMI to PSSF phases as $V$ steadily increases from an initial value
of zero and at some critical value it evolves into the PSCDW phase,
where $a$ and $b$ species of atoms reside in the opposite sides of
the lattice and each of them showing a density oscillation as
expected in the CDW phase. However, for $U=9$, the transition from
PSMI to PSCDW phase is direct with no PSSF phase sandwiched between
them. In other words for $\Delta U
> 1$ and $\Delta V >1$, each type of bosons are phase separated,
thus minimizing the energy corresponding to inter-species on-site
and nearest neighbor interactions and the PS regions
behave like a single species EBH model.

However, for $V^{ab} < V$, small value of $V$ is sufficient enough
to destroy the PSMI phase and the system evolves into the MI phase
where the densities of $a$ and $b$ bosons are equal, but with a
finite gap in the single particle energy spectrum. As $V$ increases
further the system evolves into a CDW phase with densities of both
$a$ and $b$ type atoms exhibiting oscillations. However, these
oscillations are shifted by one lattice site. This behavior is
distinctly different from the single species EBH model.

\section{Results and discussions}
We now present the details of our results. We begin with the case
$\Delta U=1.05$, $\Delta V=0.5$, $U=9$. Calculating the gap in the
energy spectrum using Eq. (\ref{eq:gap}), we observe that the system
is always gapped for the entire range of $V$. Figure
~(\ref{fig:fig1}) shows a plot of gap $G_L$ versus $1/L$ for few
values of $V$. A finite gap is a signature of the insulator phase in
the system.

\begin{figure}[htbp]
  \centering
  \epsfig{file=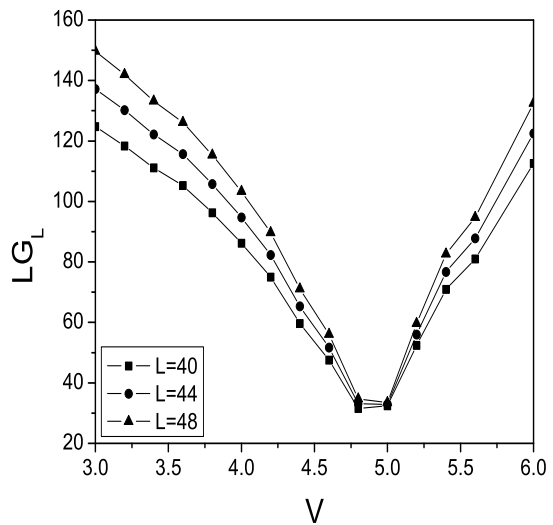,width=8cm,height=8cm}
  \caption{Scaling of gap $LG_L$ is plotted as a function of $V$ for
    different system sizes for $\Delta U=1.05$, $\Delta V=1.25$ and
    $U=9$. Gap remain finite for all the values of $V$ and shows the PSMI-PSCDW transition
    at $V_C\simeq 4.7$.}
     \label{fig:fig5}
\end{figure}

\begin{figure}[htbp]
  \centering
  \epsfig{file=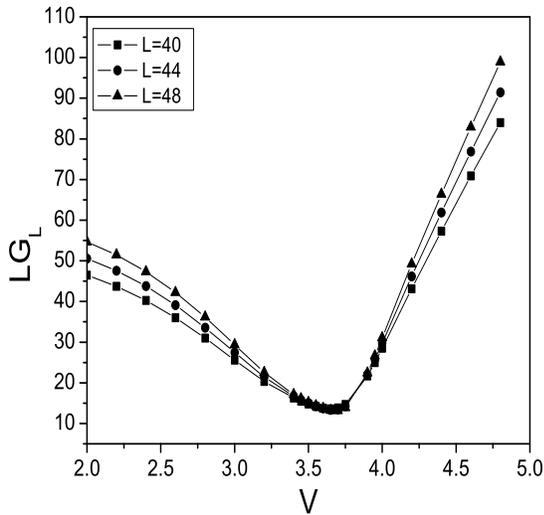,width=8cm,height=8cm}
  \caption{Scaling of gap $LG_L$ is plotted as a function of $V$ for
    different system sizes for $\Delta U=1.05$, $\Delta V=1.25$ and $U=6$.
    Coalescence of different plots between $3.4 < V < 3.9$ shows a gapless PSSF phase
    sandwiched between
    PSMI and PSCDW phases. }
     \label{fig:fig6}
\end{figure}
In order to investigate the nature of this insulator phase, we
further obtain the density distributions $\langle n_i^a\rangle$ and
$\langle n_i^b \rangle$ of both $a$ and $b$ species bosons using Eq.
(\ref{eq:ni}) and they are plotted in Fig. \ref{fig:fig2} and Fig.
\ref{fig:fig3} respectively. When V is equal to zero or very small,
the insulator phase as shown in Fig.\ref{fig:fig2} has $a$ and $b$
atoms spatially separated; i.e. it is in the PSMI phase. For small
$V$, the system behaves like a two species Bose-Hubbard model. As
$V$ increases, further the species distribute themselves through the
lattice (see Fig.\ref{fig:fig2}) thereby destroying the phase
separation. Since there is a gap in the excitation spectrum, this
corresponds to the MI phase. The critical value of $V$ for this PSMI
to MI transition is $0.2$ for $\Delta U=1.05$, $\Delta V=0.5$ and
$U=9$. Further increase of $V$ drives the system to a phase where
the two like atoms cannot occupy the adjacent sites because of large
$V$. The competition between intra- and inter species interactions
leads to an energetically favored state where the atoms arrange
themselves as shown in Fig. \ref{fig:fig3}. Both $a$ and $b$ type
bosons exhibits CDW oscillations, however, they share adjacent sites
to minimize the effect of on-site inter species interactions. The
oscillation in $\langle n^a_i\rangle$ and  $\langle n^b_i\rangle$
increases and then stabilizes at a higher $V$. This is a CDW phase
and the density oscillations of $a$ and $b$ species atoms are
shifted by one lattice site. The phase transition from MI to this
intermingled CDW phase has a critical value of $V_C \approx 1.2$,
which is obtained by plotting the CDW order parameter $O^a_{CDW}$,
for different values of $V$ ranging from $0.6$ to $3.8$ in steps of
$0.2$, versus $1/L$ as shown in Fig.~(\ref{fig:fig4}). We notice
that the $O^a_{CDW}$ goes to zero for $V < V_C\simeq 1.2$ where it
is finite for higher values of $V$. It should be noted that for the
single species extended Bose Hubbard model, the $V_C$ for MI to CDW
transition was found to be approximately equal to $4.7$ ~\cite{pai}.
Thus for $\Delta U=1.05$, $\Delta V=0.5$ and $U=9$, the nearest
neighbor interaction between the species favors a CDW over a MI
phase. The similar behavior is also seen for $U=6$. So we arrive at
the conclusion at this juncture that for $\Delta U > 1$ and $\Delta
V < 1$, the PSMI phase is unstable in the presence of a small inter
species nearest neighbor interaction. The phase diagram will then
consist of PSMI, MI and CDW phases. However, it is interesting to
note that the CDW phase is in fact  two intermingled CDW,  each for
the two different species.

\begin{figure}[htbp]
  \centering
  \epsfig{file=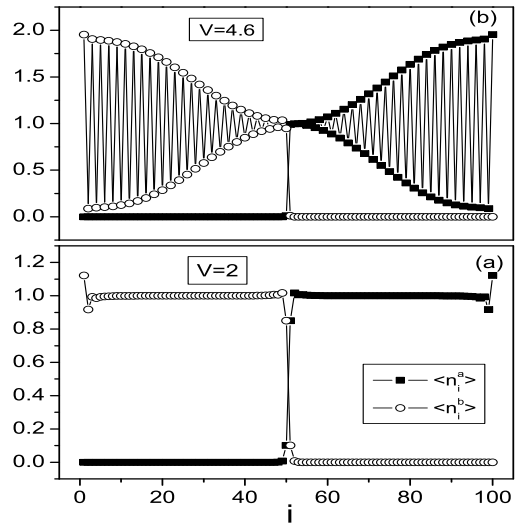,width=8cm,height=8cm}
  \caption{Plots of $\langle n^a\rangle$ and $\langle n^b\rangle$  versus $L$
  for $U=6$ and two different vales of $V$;
  (a) $V=2$ showing PSMI phase and (b) $V=4.6$ showing PSCDW phase.}
     \label{fig:fig7}
\end{figure}

\begin{figure}[htbp]
  \centering
  \epsfig{file=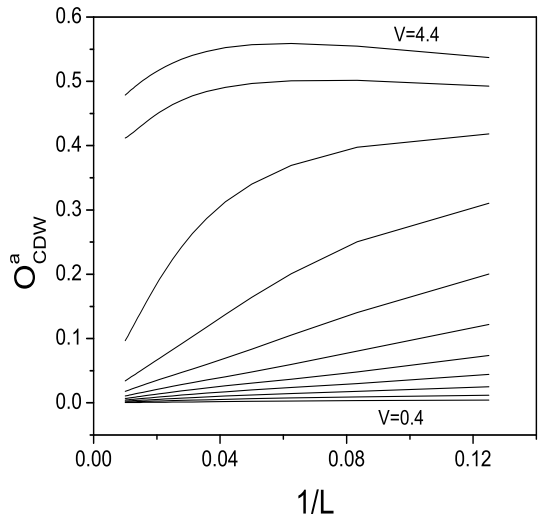,width=8cm,height=8cm}
  \caption{Plot of $O^a_{CDW}$ as a function of $1/L$
for values of $V$ ranging from $0.4$ to $4.4$ in steps of $0.4$. }
     \label{fig:fig8}
\end{figure}

Now, we proceed to discuss the other situation when $\Delta V>1$.
Considering $\Delta=1.25$, we obtain the gap $G_L$, local density
distributions ($\langle n_i^a\rangle$ and $\langle n_i^b\rangle$)
and the CDW order parameters for both $U=6$ and $9$. The most
important feature seen in this case is that the phase separation
survives for all the considered values of $V$. $a$ and $b$ species
of atoms are present in the opposite sides of the lattice. Since the
inter-species (both on-site and nearest neighbor) interactions are
larger than the intra-species interactions, the PS phase is always
energetically favored compared to the uniform case since the chances
of $a$ and $b$ atoms sharing the same site or the nearest
neighboring sites are minimized. In other words, the importance of
$U^{ab}$ and $V^{ab}$ in the present system is minimized by the PS
phase and only interactions left to compete with each other are the
on-site and nearest neighbor intra-species interactions. That means
both $a$ and $b$ atoms in the PS phase behave like a single species
EBH model. We establish these results below by analyzing the gap,
local densities and CDW order parameters.

In Fig. \ref{fig:fig5}, we plot the scaling of gap $LG_L$ as a
function of $V$ for on-site interaction $U=9$. The curves for
different lengths $L$ do not coalesce anywhere in the figure which
is the signature of the finite gap in the single particle energy
spectrum ~\cite{pai}. This implies that the phase will be either a
PSMI or a PSCDW. In contrast, different $LG_L$ curves coalesce for
$3.4 < V < 3.9$ for $U=6$ as shown in Fig. \ref{fig:fig6} suggesting
the existence of SF phase~\cite{pai} sandwiched between two gapped
phases. To understand the nature of these phases, we plot, in Fig.
\ref{fig:fig7}, $\langle n_i^a \rangle$ and $\langle n_i^b \rangle$
for two specific values of $V$, one each representing PSMI and PSCDW
phases. Phase separation can be clearly seen in these figures. Plots
of these kind yield a PSMI phase for $V < 3.4$. The phase separated
phase has the average density $\rho_a=\rho_b=1$ (see
Fig.\ref{fig:fig7}(a)). For $3.4 < V < 3.8$, the gap vanishes but
the phase separation order parameter remains finite, giving rise to
a PSSF phase. And finally for larger value of $V$, we have a clear
PSCDW phase (See Fig.\ref{fig:fig7}(b)). The CDW order parameters
plotted in Fig.\ref{fig:fig8} remain non-zero for the PSCDW phase.
It may be noted that in the PS phase when calculating the CDW order
parameter, say $O^a_{CDW}$ for $a$ bosons, only the spatially
separated regions; i.e. right hand side of the lattice is considered
since the density of $a$ bosons is zero in the left part of the
lattice. Therefore, for $U=6$, we have a transition from PSMI to
PSSF as $V$ increases. On further increase of $V$ leads to a
transition from the PSSF to the PSCDW phase. However, the transition
from PSMI to PSCDW is direct for $U=9$ as seen from
Fig.\ref{fig:fig5}. So we conclude here that for $U^{ab} > U$ and
$V^{ab} > V$, the system has a PS phase for all values of $V$ and it
behaves like a single species BH model in this PS region.

\section{Conclusions}
We have investigated the ground state properties of a two species
extended Bose-Hubbard model using the finite size density matrix
renormalization group method. We study the system for integer
filling; i.e. $\rho=\rho_a+\rho_b=1$ with $\rho_a=\rho_b=1/2$.
Starting with a phase separated Mott insulator phase (i.e. keeping
$U^{ab}>U$) and varying the nearest neighbor interaction strengths,
we predict a transition from phase separated Mott insulator to Mott
insulator and then to charge density wave phase for $V^{ab}< V$. The
charge density wave phase in this case is actually an intermingled
charge density wave phase, where both $a$ and $b$ species of atoms
show density oscillations, but are shifted by one lattice site. For
$V^{ab}< V$ the phase separation breaks for a very small nearest
neighbor interaction strength. However when $V^{ab}
> V$, the phase separation is robust. For large values of $U$, the ground
state evolves from phase separated Mott insulator to phase separated
charge density wave phase with a direct transition between them.
This is expected to be a first order phase transition ~\cite{pai}.
For smaller values of $U$, a phase separated superfluid phase is
sandwiched between the phase separated Mott insulator and phase
separated charge density wave phases. This is similar to that of a
single species extended Bose-Hubbard model except that the two
species are phase separated. We hope the present results will
stimulate future experiments.

\section{acknowledgments} We thank B. P. Das for many useful discussions. This work was
supported by DST, India (Grants No. SR/S2/CMP-0014/2007).


\begin {thebibliography}{99}

\bibitem{review}
F. Dalfovo, S. Giorgini, L. P. Pitaevskii and S. Stringari, Rev.
Mod. Phys. {\bf 71}, 463 (1999); I. Bloch, J. Dalibard, and W.
Zwerger, arXiv:0704.3011bv1 [cond-mat.other]; M. Lewenstein, A.
Sanpera, V. Ahufinger, B. Damski, A. Sen and U. Sen, Adv. Phy. {\bf
56}, 243 (2007).
\bibitem{greiner}
M Greiner, O. Mandel, T. Esslinger, T. W. HaÈnsch and I. Bloch,
Nature {\bf 415}, 39 (2002).
\bibitem{jaksch}
  D. Jaksch, C. Bruden, J. I. Cirac, C. W. Gardiner and P. Zoller
  Phys.  Rev. Lett. \textbf{81} 3108 (1998).

\bibitem{stoferle}
T. St\"{o}ferle, {\it et. al.} Phys. Rev. Lett. {\bf 92}, 130403 (2004).

\bibitem{pfau}
A. Griesmaier, {\it et. al.}, Phys. Rev. Lett. {\bf 94}, 160401 (2005).
\bibitem{pai}
R. V. Pai and R. Pandit, Phys. Rev. B {\bf 71}, 104508 (2005).
\bibitem{white}
T. D. K\"{u}hner, S. R. White and H. Monien, Phys. Rev. B {\bf 61},
12474 (2000).
\bibitem{torre}
E. G. Dalla Torre, Erez Berg and E. Altman, Phys. Rev. Lett. \textbf{97}, 260401
(2006).
\bibitem{batrouni}
G. G. Batrouni, F. H\"{o}bert and R.T. Scalettar, Phys. Rev. Lett. {\bf 97}, 087209 (2006).
\bibitem{ref1}
G. Modugno, M. Modugno, F. Riboli, G. Roati and M. Inguscio, Phys. Rev. Lett. {\bf 89}, 190404 (2002).
\bibitem{ref2}
G. Quemener, J. M. Launay and P. Honvault, Phys. Rev. A {\bf 75}, 050701 (2007).
\bibitem{ref3}
G. Ferrari, M. Inguscio, W. Jastrzebski, G. Modugno and G. Roati, Phys. Rev. Lett. {\bf 89}, 053202 (2002).
\bibitem{ref4}
T. Bourel, {\it et. al.}, Phys. Rev. Lett. {\bf 93}, 050401 (2004).
\bibitem{ref5}
T. Taglieber, A.-C. Voigt, T. Aoki, T. W. Hansch and K. Dieckmann, Phys. Rev. Lett. {\bf 100}, 010401 (2008).
\bibitem{ref6}
T. Papenbrock, A. N. Salgueiro and H. A. Weidenm\"uller, Phys. Rev. A {\bf 6}, 025603 (2002).
\bibitem{kuklov}
A. Kuklov, N. Proko\'{f}ev and B. Svistunov, Phys. Rev. Lett. {\bf 92}, 050402, (2004).
\bibitem{demler}
E. Altman, W. Hofstetter, E. Demler and M. D. Lukin, New J. Phys. {\bf 5}, 113 (2003).
\bibitem{sengupta}
A. Isacsson, M. -C. Cha, K. Sengupta and S.M. Girvin, Phys. Rev. B {\bf 72}, 184507 (2005).
\bibitem{mathey}
L. Mathey, Phys. Rev. B {\bf 75}, 144510 (2007).
\bibitem{mishra}
T. Mishra, R. V. Pai and B. P. Das, Phys. Rev. A \textbf{76}, 013604 (2007).
\bibitem{cederbaum}
O. E. Alon, A. I. Streltsov and S. Cederbaum, Phys. Rev. Lett. {\bf 97}, 230403 (2006).
\bibitem{whiteprl}
S.R. White, Phys. Rev. Lett. {\bf 69}, 2863 (1992).

\bibitem{dmrgreview}
U. Schollw\"{o}ck, Rev. Mod. Phys. {\bf 77}, 259 (2005).

\bibitem{paiprl} R. V. Pai, R. Pandit, H. R. Krishnamurthy, and S. Ramasesha, Phys. Rev.
Lett. \textbf{76}, 2937 (1996).
\end{thebibliography}
\end{document}